\documentclass[fleqn,10pt]{wlscirep}
\usepackage[utf8]{inputenc}
\usepackage[T1]{fontenc}
\usepackage{color}
\usepackage{xurl}
\usepackage{array}
\usepackage{hyperref}[colorlinks,linkcolor=blue]
\usepackage{xcolor}

\usepackage{booktabs} 
\usepackage{array}    
\usepackage{multirow} 
\usepackage{geometry} 
\title{ChineseEEG-2: An EEG Dataset for Multimodal Semantic Alignment and Neural Decoding during Reading and Listening}

\author[1a]{Sitong Chen}
\author[2a]{Beiqianyi Li}
\author[2a]{Cuilin He}
\author[1a]{Dongyang Li}
\author[1]{Mingyang Wu}
\author[1]{Xinke Shen}
\author[1]{Song Wang}
\author[3]{Xuetao Wei}
\author[4]{Xindi Wang}
\author[2*]{Haiyan Wu}
\author[1*]{Quanying Liu}

\affil[1]{Department of Biomedical Engineering, Southern University of Science and Technology, Shenzhen, China}
\affil[2]{Centre for Cognitive and Brain Sciences, Department of Psychology, Faculty of Social Sciences, University of Macau, Taipa, Macau SAR, China}
\affil[3]{Department of Computer Science and Engineering, Southern University of Science and Technology, Shenzhen, China}
\affil[4]{AI Research Institute, iFLYTEK Co., LTD, Hefei, China}

\affil[a] {Co-first authors: Sitong Chen, Beiqianyi Li, Cuilin He, Dongyang Li}
\affil[*]{Corresponding authors: Quanying Liu (liuqy@sustech.edu.cn), Haiyan Wu (haiyanwu@um.edu.mo)}

\begin{abstract}
EEG-based neural decoding requires large-scale benchmark datasets. Paired brain-language data across speaking, listening, and reading modalities are essential for aligning neural activity with the semantic representation of large language models (LLMs). However, such datasets are rare, especially for non-English languages. Here, we present ChineseEEG-2, a high-density EEG dataset designed for benchmarking neural decoding models under real-world language tasks. Building on our previous ChineseEEG dataset~\cite{mou2024chineseeeg}, which focused on silent reading, ChineseEEG-2 adds two active modalities: Reading Aloud (RA) and Passive Listening (PL), using the same Chinese corpus. EEG and audio were simultaneously recorded from four participants during $\sim$10.7 hours of reading aloud. These recordings were then played to eight other participants, collecting $\sim$21.6 hours of EEG during listening. This setup enables precise temporal and semantic alignment across the RA and PL modalities.
ChineseEEG-2 includes EEG signals, speech audio, aligned semantic embeddings from pre-trained language models, and task labels. Together with ChineseEEG, this dataset supports joint semantic alignment learning across speaking, listening, and reading. It enables benchmarking of neural decoding algorithms and promotes brain-LLM alignment under multimodal language tasks, especially in Chinese. ChineseEEG-2 provides a benchmark dataset for next-generation neural semantic decoding.

\end{abstract}

\begin{document}
\flushbottom
\maketitle
\thispagestyle{empty}

Keywords: Electroencephalogram, Brain-Computer-Interface, Brain decoding, multimodal Large language model, Chinese reading and listening

\section*{Background \& Summary}

Understanding how the human brain processes language is a longstanding challenge in both neuroscience and artificial intelligence, particularly in how linguistic information is represented across different modalities such as speech and text. Multimodal large language models (MLLM) have accelerated research in this area, showing an increasing capacity to decode natural language from brain signals~\cite{defossez2023decoding}. However, further development and validation of these models depend on the availability of large-scale datasets that provide tightly aligned neural, acoustic, and linguistic data~\cite{willett_high-performance_2023}. The current scarcity of multimodal and multilingual resources poses a fundamental challenge, as existing datasets are often constrained by a single-modality focus or limited linguistic coverage~\cite{pires2019multilingualmultilingualbert, begus_encoding_2023}.

While neural decoding models have advanced significantly, the lack of large-scale, multimodal, multilingual datasets with speech, audio, and aligned tokens has presented challenges. Multimodal LLMs such as CLAP4~\cite{elizalde2023clap}, Audiopalm~\cite{rubenstein2023audiopalm}, and Unified-IO 2~\cite{lu2024unified} have demonstrated that robust cross-modal semantic alignment is the foundation for human-inspired language modeling and neural decoding. Many of these models employ contrastive pre-training to establish a shared representational space between audio and text, which in turn serves as a target for mapping neural signals in brain-to-language alignment. Recent approaches like BELT-2~\cite{zhou2024belt2bootstrappingeegtolanguagerepresentation} and Thought2Text~\cite{mishra2025thought2texttextgenerationeeg} focus on tasks such as fine-grained brain-to-language alignment, sequential decoding of neural signals, and large-scale cross-lingual or cross-modal knowledge transfer, which rely on extensive token-level aligned multimodal datasets that precisely match brain, speech, and text representations. However, datasets that can support this alignment are not yet widely available.

Despite progress in unimodal EEG corpora, existing datasets fall short of the requirements set by recent LLM architectures. Most provide only single-modality recordings (e.g., silent reading in ZuCo~\cite{hollenstein2018zuco, hollenstein2019zuco} or auditory listening in Broderick et al.~\cite{broderick2018electrophysiological}, Brennan et al.~\cite{brennan2019hierarchical}), lacking the instance-level correspondence of linguistic content across modalities required for effective cross-modal and multi-task neural learning. Moreover, datasets such as those from Schoffelen et al.~\cite{schoffelen2019204} and Gwilliams et al.~\cite{gwilliams2023introducing} leverage MEG but lack both task diversity and scalable modality coverage. More recently developed multimodal datasets, including CineBrain~\cite{gao2025cinebrainlargescalemultimodalbrain}, Natural Scenes Dataset (NSD)~\cite{allen_massive_2022}, and Narratives dataset~\cite{nastase_narratives_2021}, offer expanded modality coverage but lack precise token-level language alignment or comprehensive multimodal semantic correspondence necessary for speech-focused modeling. Furthermore, nearly all these existing resources are limited to English, leaving a pronounced gap for Chinese and other non-English languages, restricting the development of next-generation neural decoding, brain-inspired language modeling, and the multilingual advancement of BCI applications.

To address this limitation, we introduce ChineseEEG-2: a multimodal EEG dataset covering both RA and PL tasks. Using the same linguistic corpus as ChineseEEG, we recorded EEG and audio data from four participants for approximately 10.7 hours of reading aloud. Their recordings were then played to eight additional participants in a listening task, resulting in 21.6 hours of EEG data aligned to the same semantic content.With fine-grained temporal and linguistic alignment, ChineseEEG-2 is the first large-scale token-aligned Chinese EEG benchmark supporting comparative studies across reading, speaking, and listening modalities. Together with the original ChineseEEG dataset, ChineseEEG-2 extends token-level semantic alignment to include silent reading, speaking, and listening, all based on the same naturalistic corpus, offering researchers new opportunities for cross-modal neural decoding studies in Chinese language processing.

In summary, ChineseEEG-2 is designed to serve as a benchmark dataset for semantic neural decoding. Through providing paired EEG-speech recordings, cross-modal temporal precision, and language-level alignment in a non-English context, ChineseEEG-2 overcomes longstanding barriers of language and modality, making it an essential resource for cross-modal research and for non-English brain-language decoding. With its ability to align semantic information from brain activity and LLMs, ChineseEEG-2 supports the evolution of the next generation of brain-computer interface (BCI) systems. It opens new possibilities for recent LLMs-driven BCI paradigms, such as instance-wise cross-modality mapping, sequential decoding, and multimodal fusion, and helps close the gap between human brain representations and computational semantic models. With its rich multimodal structure and matched linguistic stimuli, ChineseEEG-2 contributes a scalable, culturally diverse resource to the broader goal of human-level language decoding from brain activity. Building on the original ChineseEEG dataset, ChineseEEG-2 broadens the possibilities for research. Together, our resources support three main research directions:
\begin{itemize}
    \item \textbf{Cross-modal neural decoding}, enabling models to generalize across sensory input and output modalities;
    \item \textbf{Brain-to-MLLM alignment}, grounding shared semantic representations learned by language-audio models in real neural data;
    \item \textbf{Standardized benchmarking}, enabling fair comparisons across decoding models, fusion strategies, and cross-modality training methods.
\end{itemize}

\section*{Methods}



\subsection*{Participants and Task Overview}

A total of 12 healthy participants were recruited for the study, ranging in age from 18 to 25 years (mean age: 21.9 years; 4 males, 8 females). All participants were right-handed, had normal or corrected normal vision, and had normal hearing. No participant had a self-reported history of neurological or psychiatric disorders. Each participant voluntarily participated in the study and gave their informed consent before the experiment. In the RA task, each participant completed 4 experimental runs. For the PL task, all participants completed 4 runs except participant 4, who experienced an interruption during the second run and therefore completed 5 runs in total. They received monetary compensation in the form of approximately 50 MOP coupons (Macanese Pataca, the official currency of the Macao Special Administrative Region of China) for each experimental run.  This study complied with the Declaration of Helsinki and was approved by the Institutional Review Board of the University of Macau (Approval No. $BSERE20-APP011-ICI$).

Before the formal experiment, all participants underwent a pre-experimental session to validate the experimental procedures and ensure that the time triggers were synchronized in the audio and EEG recordings. Among the 12 participants, four (2 males and 2 females) conducted the RA task, while the remaining eight conducted the PL task. In the formal experiment, in total, 10.8 hours of reading data (around 3 hours per participant) and approximately 21.6 hours of listening data (around 3 hours per participant) were collected, a total of 32.4 hours of data. The EEG data of participant 08 during the PL task had a missing 10-minute trigger in run 2 of \textit{The Little Prince}. These incidents were documented to account for any potential biases or data gaps during the analysis phase.

\subsection*{Experimental Materials} 

The experimental materials comprised the Chinese translation of Antoine de Saint-Exupéry’s \emph{The Little Prince} and excerpts from Shen Shixi’s  \emph{Garnett Dream}. These texts were divided into uniform-length passages to standardize reading sessions and facilitate subsequent analysis. The choice of these works as semantic stimuli was primarily driven by the need to maintain compatibility with the existing ChineseEEG dataset~\cite{mou2024chineseeeg}, ensuring consistency and avoiding discrepancies that could arise from heterogeneous materials. Both texts are exemplary pieces of children's literature, written in accessible language that predominantly utilizes high-frequency vocabulary, thereby minimizing cognitive load during comprehension. Their rich linguistic diversity and varied syntactic structures provide an ideal foundation for in-depth linguistic and semantic analyses, as well as cross-modal comparisons between reading and auditory processing. Importantly, \emph{The Little Prince} is recognized as a classic in Western literature, while \emph{Garnett Dream} holds significant cultural and educational value in Chinese children's literature.

In the experiment, the selected texts were divided into multiple sections. Specifically, \emph{The Little Prince} consists of 27 chapters, encompassing the complete narrative, while \emph{Garnett Dream} was divided into 9 sections for participants f1, f2, and m2 in the RA task and participants 01-04 in the PL task, and 8 longer sections for participant m1 in the RA task and participants 05-08 in the PL task. The arrangement of the \emph{Garnett Dream} audio recording sessions was adjusted for participant m1 (male) in the reading aloud task, with one section less content but more content (8 sections in total) compared to participant f1 (female). Therefore, participants 05-08 in the PL task provided more EEG data in fewer sections, corresponding to the segments provided by participant m1. To ensure equivalent session durations and control for potential confounding variables related to exposure time, the length of \emph{Garnett Dream} sections was adjusted to match the total length of \emph{The Little Prince}.
In general, the text materials we used consisted of 46,591 Chinese characters, 21,829 characters from \emph{The Little Prince} and 24,762  characters from \emph{Garnett Dream}.

\subsection*{Experimental procedures} 
Participants were seated in an ergonomic chair positioned around 67 cm from a $54 cm \times 30.375 cm$ LCD monitor ($1920 \times 1080$ resolution, 60 Hz refresh rate), replicating the ChineseEEG experimental conditions (Figure~\ref{fig:pipeline}a). The stimuli consisted of two Mandarin novels: The Little Prince and Garnett Dream,  divided as in the Experimental Materials section, segmented to balance duration (mean run time = $25 \pm 3$ min). The introduction passages served as practice trials to familiarize participants with the task requirements and minimize potential learning effects.



\subsubsection*{Reading-aloud task}

During the RA task, participants were presented with three scrolling lines of text and instructed to read aloud the centrally displayed red-highlighted line (Figure~\ref{fig:pipeline}b), which was implemented using PsychoPy v2023.2.3\cite{Peirce2019PsychoPy2EI}. The text was presented at a rate of 0.25 seconds per character, calibrated to approximate natural reading speed while ensuring clear articulation. The materials were segmented with natural dialogue markers (e.g., quotation marks) and syntactic boundaries to preserve semantic coherence. Each trial was followed by a mandatory 20-second interval, with participants controlling progression via spacebar press, and extended one-hour breaks were provided between novels to prevent fatigue.

The reading materials were systematically divided into segments based on both content structure and temporal considerations. The Little Prince was organized into 27 content-based segments, while Garnett Dream was partitioned into 10-minute temporal segments, each preceded by a 5-second silent period to facilitate attentional preparation. This segmentation strategy allows for discrete analysis of cognitive processing while maintaining natural reading flow. The complete experimental protocol included one practice session followed by two formal reading sessions per participant, with each session divided into two experimental runs (Session 1: Run 1 - Ch.1-14, Run 2 - Ch.15-27 of The Little Prince; Session 1: Run 1 - Ch.1-5, Run 2 - Ch.6-9 of Garnett Dream for participant f1; Run 1 - Ch.1-4, Run 2 - Ch.5-8 of Garnett Dream for participant m1; where Ch. denotes manually segmented chapters). The session durations were capped at two hours to prevent fatigue-related performance degradation. 

High-fidelity audio recordings were collected throughout the reading sessions using standardized procedures. These recordings underwent post-processing using Auphonic and Audacity software for noise reduction and quality optimization. For subsequent experimental phases, gender-balanced stimulus sets were created by selecting recordings of one male and one female participant based on objective acoustic quality metrics, including clarity and articulation precision.

The experimental design incorporated multiple safeguards to ensure data quality and participant comfort. The duration of the sessions was capped at 2 hours to prevent fatigue, and the self-paced progression system allowed participants to extend rest periods as needed. The controlled presentation parameters provide a reliable assessment of the reading processes while maintaining participant engagement throughout the study.

\subsubsection*{Passive-listening task}

The PL task was conducted at the University of Macau under visual display conditions identical to those of the reading task. The PL task utilized processed audio recordings from the reading-aloud phase as experimental stimuli. Two representative voice samples (one male and one female) were selected based on objective acoustic parameters, including pronunciation clarity, speech fluency, and recording quality, to serve as standardized auditory stimuli. Eight participants were randomly assigned to the male-voice or female-voice condition (n = 4 per group). The stimuli were systematically organized into four folders per voice condition, containing chronologically segmented audio files from both literary works (Run 1 - Ch.1-14, Run 2 - Ch.15-27 of The Little Prince; Session 1: Run 1 - Ch.1-5, Run 2 - Ch.6-9 of Garnett Dream for participants 01-04, Run 1 - Ch.1-4, Run 2 - Ch.5-8 of Garnett Dream for participants 05-08, where Ch. denotes manually segmented chapters) to ensure balanced exposure across experimental sessions.

\begin{figure}[ht]
\centering
\includegraphics[width=\linewidth]{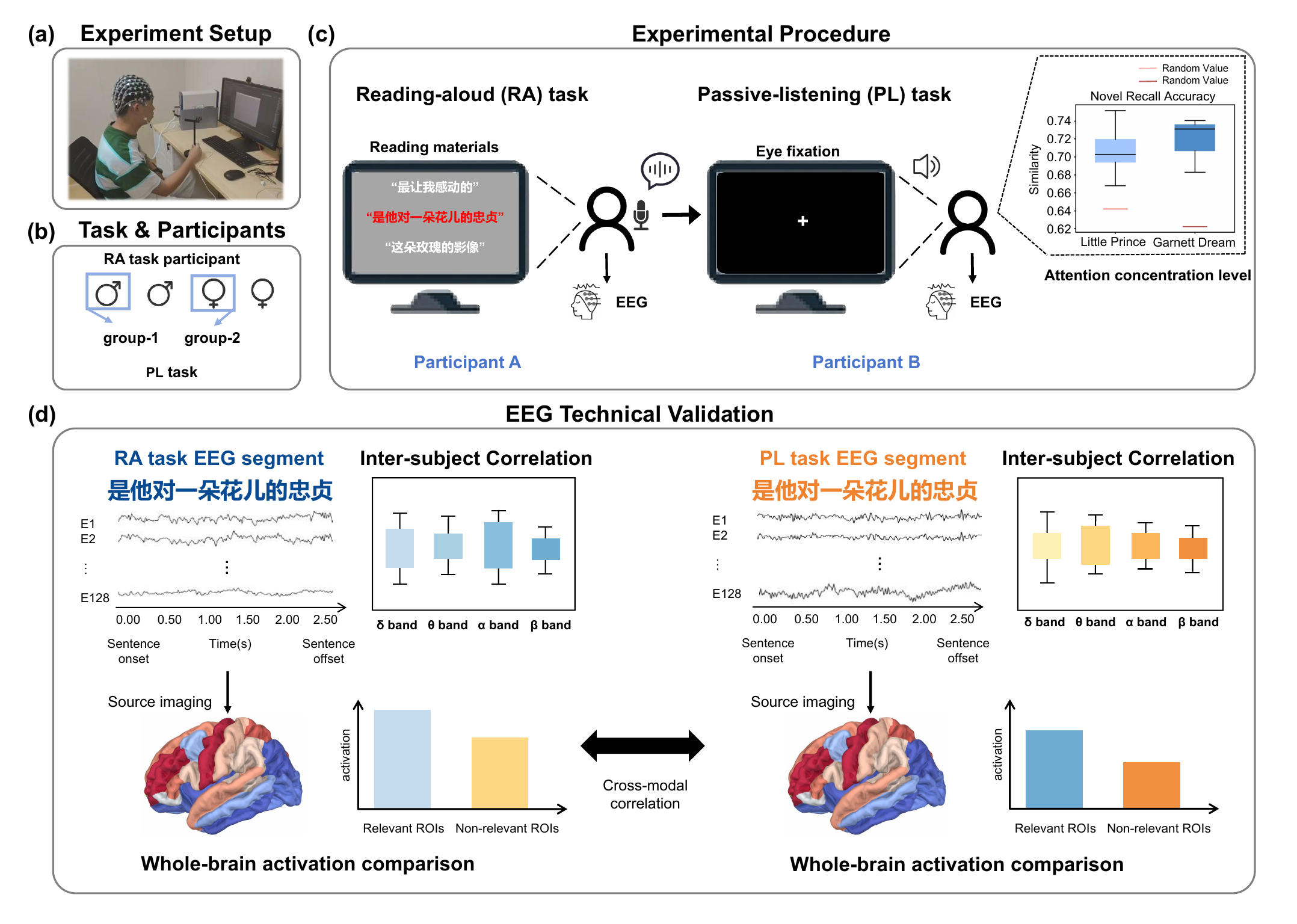}
\caption{Overview of the experiment and the modalities included in the dataset. (a) The experiment setup.  Participants were instructed to sit quietly approximately 67cm from the screen and sequentially read the highlighted text. (b) The subject \textcolor{red}{arrangement, only} the audio from female and male subject 1 in Reading-aloud group were chosen as materials for the two groups of subjects in Passive-listening task.  (c) The experimental \textcolor{red}{paradigm}. Participants' 128-channel EEG signals were recorded while reading the highlighted sentence or listening to the audio.  (d) Data alignment in \textcolor{red}{ChineseEEG-2}. The corresponding audio and EEG segment to the same sentence are converted to embeddings and \textcolor{red}{semantically} aligned.}
\label{fig:pipeline}
\end{figure}

During stimulus presentation, participants maintained visual fixation on a central white cross to minimize ocular artifacts in EEG recordings (Figure~\ref{fig:pipeline}b). The protocol incorporated structured breaks including mandatory 10-second inter-segment intervals and self-paced rest periods between experimental runs. Following each segment, participants completed attention monitoring questionnaires that contained both content-specific comprehension questions and subjective attentiveness ratings. Post-session verbal recall tasks were administered to assess memory retention and narrative comprehension.

Task engagement was objectively validated through computational linguistic analysis comparing Whisper-large-v3-generated transcripts of participants' oral recalls with original story embeddings. The significantly higher semantic similarity between participant recalls and original texts compared to randomly generated narratives (Figure~\ref{fig:pipeline}c) confirmed appropriate cognitive engagement with the auditory stimuli throughout the experimental sessions. This quantitative validation complemented behavioral measures to ensure data quality and task compliance.

\subsection*{EEG Data collection and analysis}

\subsubsection*{EEG data collection}

EEG signals were recorded using a 128-channel HydroCel Geodesic Sensor Net from Electrical Geodesics Inc., providing full scalp coverage.
Dual acquisition protocols were implemented for both tasks. During the RA task, EEG data were sampled at 250 Hz with sentence-level triggers marking the onset and offset of each text line, synchronized to the presentation rate of 0.25 s/character. For PL sessions, the sampling rate was increased to 1000 Hz. To mitigate the risk of data loss resulting from issues such as trigger misfires or power outages, the EEG recordings during the PL task were divided into two to three segments. The triggers indicated chapter sequences relative to the start of segments, as opposed to the entirety of the novel, which was the case during the reading phase.
Signal quality was assured by pre-experimental impedance checks, which maintained all electrodes below 50 k$\Omega$ through saline solution replenishment. The raw EEG data were exported to metafile format (.mff) files on the macOS system, preserving the task-specific data (\texttt{task-reading / lis}) and session identifiers (\texttt{ses-littleprince/ses-garnettdream}).

\subsubsection*{EEG data pre-processing}

\begin{figure}[ht]
\centering
\includegraphics[width=\linewidth]{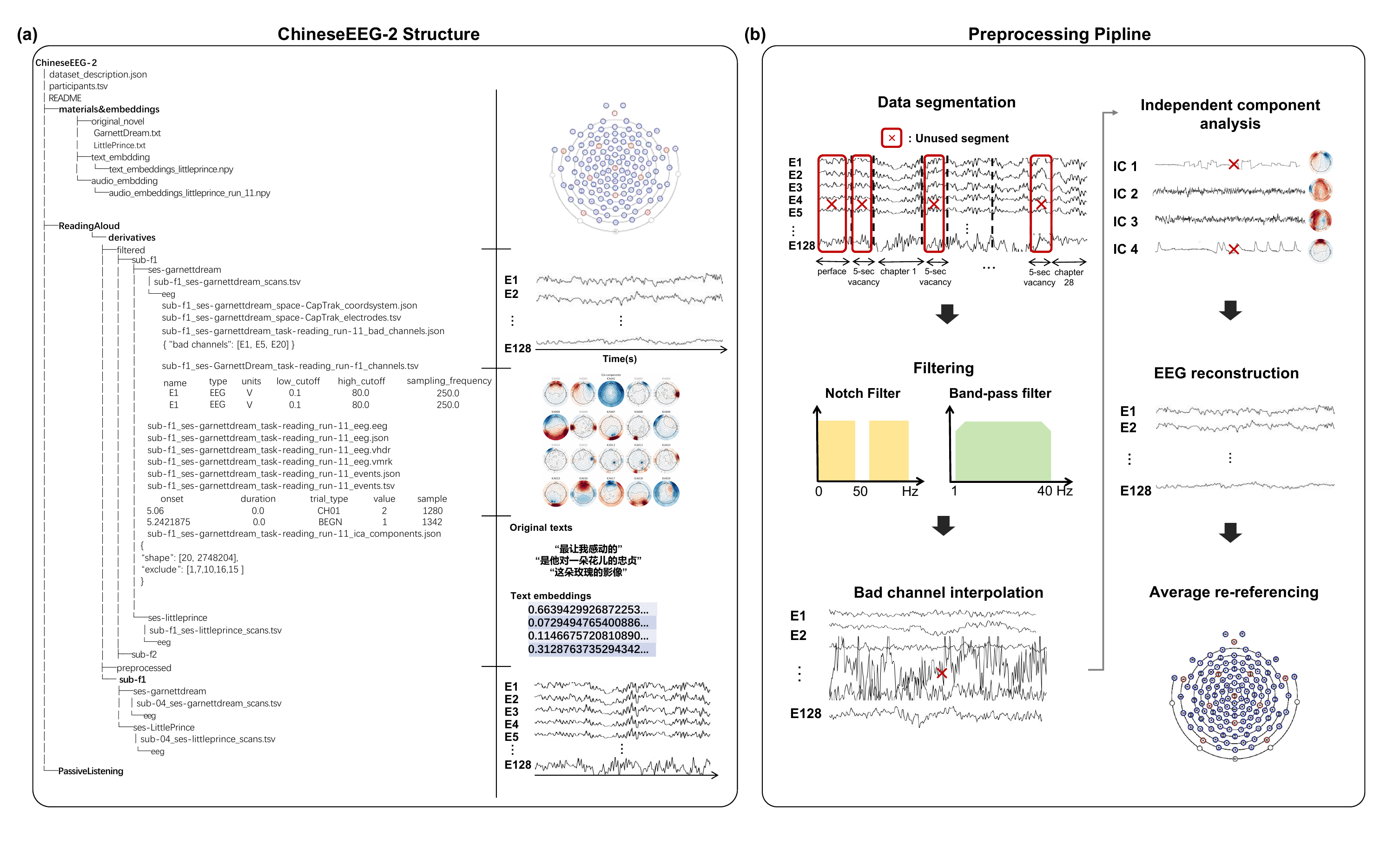}
\caption{\textcolor{red}{Directory structure of the ChineseEEG-2 dataset and EEG preprocessing pipeline. (a) File structure of the dataset. The file structure encompasses experiment metadata, detailed electrode information (including location, type, and sampling rate), annotations for channels identified as defective during pre-processing, EEG data in BrainVision format, event files containing marker information, records of removed independent components during pre-processing, as well as the original text and embedding files. (b) EEG pre-processing pipeline. The pipeline includes data segmentation, band-pass filtering, interpolation of faulty channels, independent component analysis (ICA) for denoising, and re-referencing.}}
\label{fig:modality}
\end{figure}

The EEG preprocessing pipeline implemented through MNE-Python v1.6.0 (Figure~\ref{fig:modality}b) followed a neurophysiologically-grounded sequence to ensure signal integrity for language-related analyses. The conversion to BrainVision format was performed using the MNE-BIDS v0.14 Python package to ensure complete compliance with the BIDS specification. Initial data trimming removed non-task intervals, including practice trials from both novels' prefaces and 5-second silent buffers preceding each chapter. The retained epochs were downsampled to 250 Hz, preserving temporal resolution while reducing computational load.
Powerline interference was mitigated via a 50 Hz notch filter, followed by 1-40 Hz bandpass filtering to isolate neurocognitive rhythms: delta (0.5-4 Hz) for sustained attention, theta (4-8 Hz) for semantic integration, alpha (8-12 Hz) for inhibitory control, and beta (12-30 Hz) for predictive processing, targeting the frequency bands most relevant to cognitive processing. Channels exhibiting excessive noise were reconstructed by spherical spline interpolation, preserving topographic continuity across the 128-channel montage.
Artifact rejection employed ICA decomposition (FastICA algorithm, 40 components) with semi-automated classification: ocular components identified via frontal scalp topography and high kurtosis, cardiac artifacts through periodic peaks. For details of the ICA components for each participant, please refer to \url{https://github.com/ncclab-sustech/ChineseEEG-2/blob/main/data_preprocessing/README.md}. Residual DC shifts and slow drift artifacts were removed through baseline correction and average re-referencing, normalizing signals across participants for group-level analyses. The fully processed data were stored in BIDS-derivatives format, ready for sensor-space and source-level investigations of semantic processing dynamics.

\subsection*{Temporal Alignment of EEG, Text, and Audio Sequences} 

We synchronized the EEG data and the corresponding text or audio stimuli using a temporal synchronization protocol. Trigger alignment was implemented through: (1) hardware-level trigger synchronization and (2) character-level refinement of temporal markers (Figure \ref{fig:pipeline}d). During data acquisition in both RA and PL tasks, triggers were embedded in the EEG recordings time-locked to the onset and offset of each presented stimulus, marking the exact start and end of every text line. These triggers enabled initial alignment of EEG recordings with the stimulus. Character-level temporal mapping was achieved using the fixed presentation rate of 0.25 seconds per character. In the case of textual stimuli, temporal alignment was achieved through a direct comparison of individual character timestamps with the corresponding markers in the EEG data. For auditory stimuli, temporal localization of each character was accomplished by determining its onset time and performing synchronized identification across both the EEG recordings and the audio signal.

\section*{Data Records}

The data modalities included in the structure of the ChineseEEG-2 dataset are shown in Figure \ref{fig:modality}a, including raw data and derivatives. Raw data contains raw EEG, raw text, and audio materials, and derivatives contain pre-processed EEG data, audio and text embeddings generated by model Wav2Vec2\cite{enrique_hernández_calabrés_2024} and Bert-base Chinese~\cite{devlin2018bert}.

The complete dataset is publicly accessible at \url{https://doi.org/10.57760/sciencedb.CHNNeuro.00001}.

\subsection*{Data organization}
The dataset adheres to the EEG-BIDS (Brain Imaging Data Structure) specification, which extends the standard BIDS framework for electroencephalography (EEG) data. As depicted in Figure~\ref{fig:modality}a, under the root directory are the folders\textit{ Reading aloud }and\textit{Passive Listening}, respectively containing the EEG data in the reading aloud task and in the listening task. The data of each participant is encapsulated in a compressed archive (.zip), containing standardized EEG file formats: \textit{.vhdr} (header), \textit{.vmrk} (marker), and \textit{.eeg} (raw signal). The EEG data are organized into two session-specific subfolders, \textit{ses-littleprince} and \textit{ses-garnettdream}, corresponding to the two semantic materials. The EEG data for all participants are separated by chapter for efficient processing and analysis. To facilitate analysis, the EEG recordings are segmented with run numbers encoded as \textit{run\-$XY$}, where \textit{X} is the first number and denotes the run index, \textit{Y} is the chapter sequence within that run (e.g., \textit{run\-112} = Chapter 12 in Run 1; \textit{run\-21} = Chapter 15 in Run 2). The \textit{derivatives} folder contains processed EEG data and filtering results, all structured the same as the raw data. The \textit{filtered} folder contains band-pass filtered EEG data within the frequency range (1-40 Hz). These filtered data files enable focused analyses on different neural oscillatory activities associated with linguistic tasks. Each participant has an individual subfolder located under \textit{preprocessed}, which contains EEG data that have undergone our preprocessing pipeline. The \textit{materials\&embeddings}  are provided in the RA task folder, which stores the text stimuli used during the experiment and generated embeddings. Original novels are preserved in \textit{.txt} format, while embeddings of audio and text materials are preserved in .npy files. 

The dataset is organized according to EEG-BIDS guidelines to ensure reproducibility, transparency, and ease of use for other researchers. It is valid for integration with various analytical pipelines and offers a robust foundation for more advanced research in EEG-based language studies.

\section*{Technical Validation}

\subsection*{Inter-subject correlation analysis}

\begin{figure}[ht]
\centering
\includegraphics[width=\linewidth]{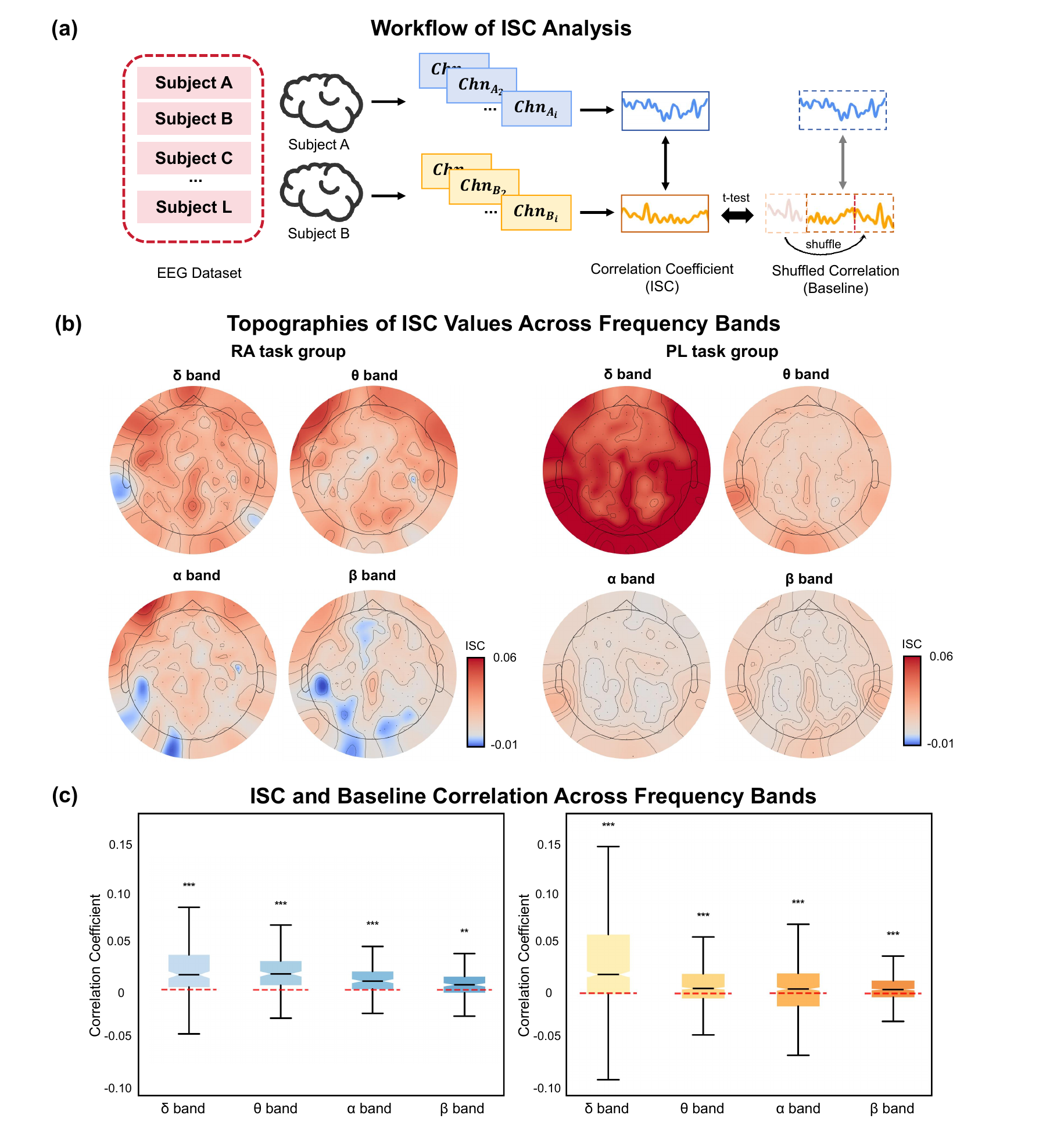}
\caption{\textcolor{red}{Inter-subject correlation analysis of EEG data. (a) Schematic illustration of the validation procedure comparing electrode-wise correlation coefficients between pairs of subjects, with statistical testing against a shuffled baseline. (b) Topographic maps showing ISC coefficients across electrodes for the $\delta$, $\theta$, $\alpha$, and $\beta$ frequency bands during the RA and PL tasks. Warmer colors indicate higher ISC values. (c) Boxplots of correlation coefficients for each frequency band compared to the shuffled baseline, demonstrating significantly higher actual correlations (One-sample t-test, $**p<0.01$, $***p<0.001$), validating the neural plausibility and authenticity of the EEG signals.}}
\label{fig:analysis}
\end{figure}

To assess whether the recorded EEG signals reflect neural responses to the linguistic stimuli shared across subjects, we performed inter-subject correlation (ISC) analysis at the electrode level across four canonical frequency bands: delta, theta, alpha, and beta. For each frequency band and each subject $i$, we computed the correlation of each pair of electrodes between subject $i$ and all other subjects $k$, where $k \in \{1, \dots, n\}$ and $k \neq i$. $n$ is the total number of subjects (Figure~\ref{fig:analysis}a). To determine whether the observed correlations were significantly above chance, we constructed the baseline correlation using a circular time-shift shuffle procedure~\cite{Parra_2019}. Specifically, for each electrode, the EEG time series was segmented at a random time point and temporally shuffled by concatenating the latter segment before the former, thereby disrupting temporal alignment while preserving spectral properties. This procedure was repeated 500 times for each subject pair and electrode, and the average correlation coefficients across all iterations, electrodes, and subject pairs were used as the baseline (Figure~\ref{fig:analysis}a).

Figure~\ref{fig:analysis}b presents the scalp topography of the resulting correlation coefficients, highlighting consistent spatial patterns across subjects. Figure~\ref{fig:analysis}c shows the boxplot of ISC from each band. Paired-sample t-tests revealed that authentic ISC values were significantly higher than the baseline correlation in all frequency bands (One-sample t-test, $ps<0.001$), confirming that EEG signals contain neural information related to shared stimuli, supporting the validity and quality of the dataset.


\subsection*{EEG Source localization and cross-modal analysis}

The 128-channel EEG data supported source localization analyses. All source-level analyses were performed with MNE-Python, following this pipeline: (1) Cortical surface reconstruction using the fsaverage MRI template \cite{Fischl1999HighresolutionIA}; (2) Source space construction (10,242 sources/hemisphere) via Connectome Mapper 3 \cite{TourbierJOSS2022}; (3) Implementation of a 3-layer Boundary Element Method (BEM) model (15,360 triangles; conductivities: brain=0.3 S/m, skull=0.006 S/m, scalp=0.3 S/m) following the ChineseEEG protocol \cite{mou2024chineseeeg}i; and (4) Reconstruction of EEG sources using dynamic Statistical Parametric Mapping (dSPM), selected for its established reliability in EEG source imaging. 

The visualization of the source activities corresponding to each word in the example sentence is provided. The result of source localization demonstrates a more focused activation near the left middle temporal gyrus in subjects in the RA task, which is related to language comprehension\cite{pobric2009role}. For subjects in the PL task, it is observed that activation areas are more dispersed. Activation areas included not only the anterior temporal lobe and the temporal-parietal region, which are associated with language comprehension and primary processing\cite{bi2011role}, but also the prefrontal region, which is associated with visual processing. The potential cause may be that the subjects in the PL task are less focused, as there is less engagement compared to the RA task. Additionally, participant task engagement  in PL task was guaranteed by measuring the similarity of the embeddings of the transcript of the collected audio where the subjects recalled the story in their own words and the embedding of the original story. The transcripts were generated using the whisper-large-v3 model\cite{radford2022whisper}. For both semantic materials, the average similarity of the audio of the subjects is above the story randomly generated by the natural language model using frequently appearing words as in Figure \ref{fig:pipeline}c. In general, the result shows that ChineseEEG-2 contains data where neural activations respond to the given semantic materials .

To verify task-induced activation of language networks, we conducted source analysis in language-related regions guided by Lausanne2018.scale2\cite{CAMMOUN2012386}, with language-specific ROIs identified per Bizley's auditory-language mapping \cite{Bizley2013}. For details, please refer to \url{https://github.com/ncclab-sustech/ChineseEEG-2/blob/main/data_analysis/README.md}.  The analysis focused on beta-band oscillatory activity (14-30 Hz), extracted through the Hilbert transform of bandpass-filtered signals.

\begin{figure}[ht]
\centering
\includegraphics[width=\linewidth]{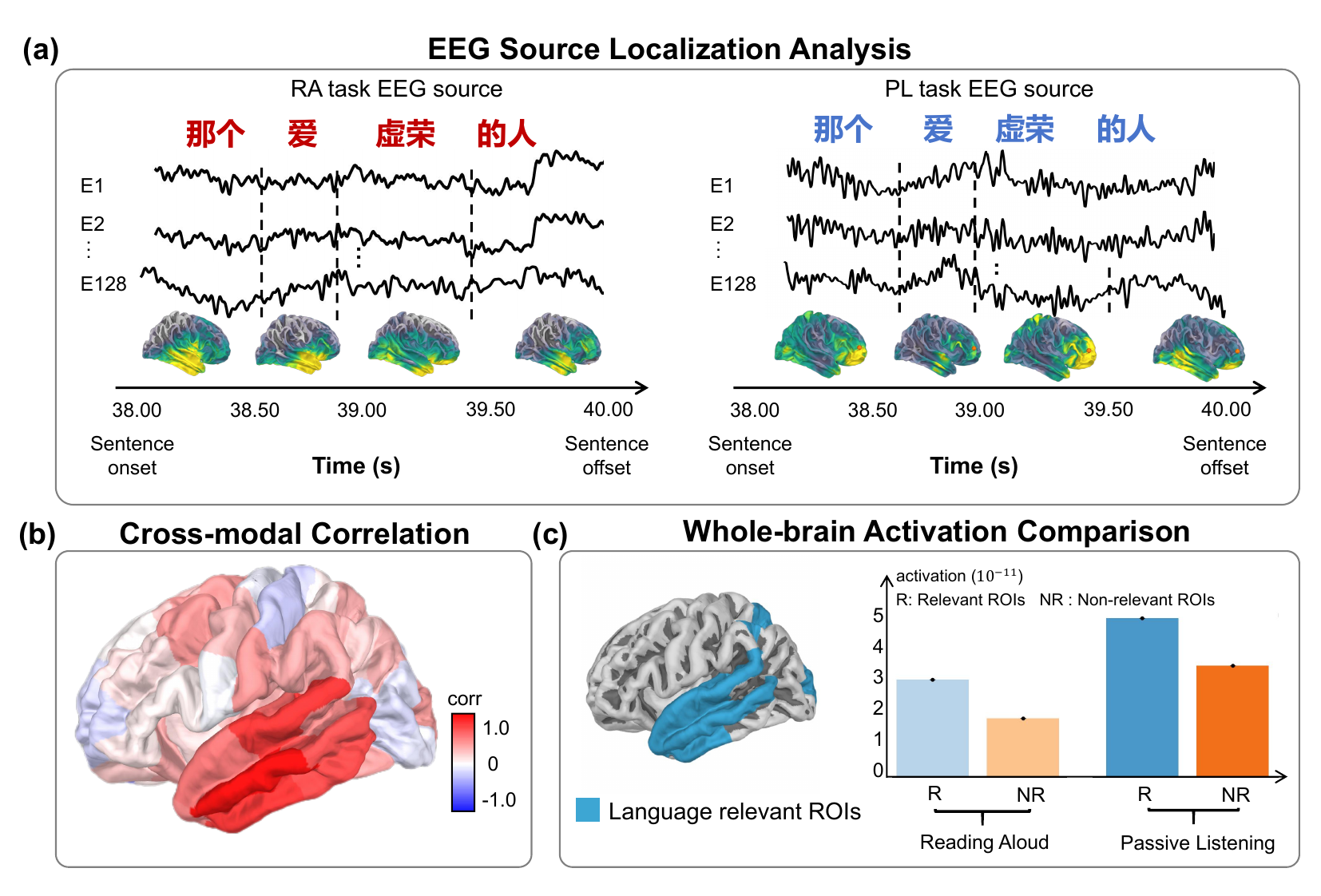}
\caption{\textcolor{red}{EEG Source-level analysis. (a) Source localization during the reading-aloud task. EEG sensor-level data from the reading-aloud task were localized using dynamic Statistical Parametric Mapping (dSPM)~\cite{dale2000dynamic}, consistent with ChineseEEG dataset~\cite{mou2024chineseeeg}. The reconstructed source maps corresponding to the time window marked by the highlighted sentence were presented.  (b) Cross-modal correlation analysis. The correlation coefficients between the neural representations in reading-aloud task and passive-listening task were illustrated. (c) Task-dependent activation of language regions. Average activation levels were computed for language-relevant regions (blue) and control regions (orange) across multiple subjects in both the RA and PL tasks. Activation in language regions was significantly higher during the reading-aloud task compared to control regions (Welch's t-test, $p < 0.001$).}}
\label{fig:source}
\end{figure}

Cross-modal correlation analysis was performed by computing the pairwise similarity of amplitude envelopes between the RA and PL task groups at each cortical parcel. For each group, individual amplitude envelopes were averaged within groups to obtain representative group-level signals, then pairwise correlations for each brain region were calculated across the two task groups. As shown in Figure~\ref{fig:source}b, the left superior temporal gyrus and the left middle temporal gyrus, which are both related to language processing, exhibited stronger cross-task correlations than between the left superior temporal gyrus and the left pericalcarine cortex, a primary sensory area\cite{10.1162/jocn_a_00244}(Steiger’s Z-test, $p < 0.05$). The result was obtained after excluding two participants (03, 04) with excessive artifacts. The correlation between task-activated regions substantiates the dataset's sensitivity to language-related neural dynamics.


\section*{Usage Notes}


The code for the experiment and data analysis has been uploaded to GitHub to facilitate sharing and use, accessible at \url{https://github.com/ncclab-sustech/ListeningEEG}.

The code repository contains five main modules, including scripts for preparing the materials, reproducing the experiment, data processing, data embedding, and data analysis procedures. The script \texttt{cut\_chinese\_novel.py} in the \texttt{novel\_segmentation} folder contains the code to prepare the stimulation materials from the source materials. Both files \emph{littleprince.xlsx} and \emph{garnettdream.xlsx} record the sections of the materials, with \emph{garnettdream.xlsx} containing the end of each segment for participants f1 and m1 in the RA task. The script \texttt{ui\_experiment.py} contains code for the RA task experiment and the \texttt{recording.py} enables trigger sending, which should be guaranteed to be initiated first in the experiment. The script \texttt{experiment\_listen.py} contains code for the PL task experiment. The script \texttt{preprocessing.py} in \texttt{data\_preprocessing} folder contains the main part of the code to apply pre-processing on EEG data and conversion to BIDS format. The script \texttt{text\_embed.py} in the \texttt{embeddings}  folder contains code to generate embeddings for semantic materials, while the script \texttt{audio\_embed.py} is for generating audio embeddings. The script \texttt{isc\_analysis.py} contains necessary code for ISC analysis, and the script \texttt{source\_analysis.ipynb} is for calculating forward and inverse solutions to reconstruct the source space. The code for EEG data pre-processing is highly configurable, permitting flexible adjustments of various pre-processing parameters, such as data segmentation range, downsampling rate, filtering range, and choice of ICA algorithm, thereby ensuring convenience and efficiency. Researchers can modify and optimize this code according to their specific requirements.

Before using our ChineseEEG-2 dataset \url{https://doi.org/10.57760/sciencedb.CHNNeuro.00001}, we encourage all users to check \textit{README.md} and the updated information in the GitHub repository.

\section*{Code availability}
The code for all modules is openly available at \url{https://github.com/ncclab-sustech/ChineseEEG-2}. 
All scripts were developed in Python 3.10\cite{10.5555/1593511}. Package openpyxl v3.1.2 was used to export segmented text in Excel (.xlsx) files, egi-pynetstation v1.0.1, and psychopy v2022.2.4\cite{Peirce2019PsychoPy2EI} were used to implement the scripts for EGI device control and stimuli presentation. In the data pre-processing scripts, MNE v1.8.0\cite{GramfortEtAl2014} and pybv v0.7.5\cite{Appelhoff_pybv_--_A}were used to implement the pre-processing pipeline, while mne-bids v0.14~\cite{dd3ae731a9a34ae8b9e94ebe5debc83d, pernet2019eeg} was used to organize the data into BIDS\cite{Gorgolewski2016, pernet2019eeg} format. The audio and text embeddings were calculated using Hugging Face transformers v4.36.2\cite{wolf-etal-2020-transformers}.
For more details on code usage, please refer to the GitHub repository.

\bibliography{reference}


\section*{Acknowledgements}
This work was mainly supported by the National Key R\&D Program of China (2021YFF1200804), Science and Technology Development Fund (FDCT) of Macau [0127/2020/A3, 0041/2022/A], the Natural Science Foundation of Guangdong Province(2021A1515012509), and the SRG of the University of Macau (SRG2020-00027-ICI). We also thank Dr Qingqing Qu for the suggestion on the design, all of the collaborators for the ChineseEEG dataset, and all participants involved in ChineseEEG-2.

\section*{Author contributions statement}
Q.Liu and H.Wu designed the study; S.Chen, D.Li wrote the code for experiments; S.Chen conducted the reading aloud experiments, B.Li, C.He conducted the listening experiments, S.Chen analyzed the results.  S.Chen wrote the first draft. All authors checked the code, wrote the manuscript, reviewed the manuscript, and approved the final manuscript.

\section*{Competing interests}
The authors declare no competing interests.

\begin{table}[!htbp]
\centering
\caption{Overview of experiment sessions} 
\small 
\begin{tabular}{
  c 
  c 
  c 
  >{\centering\arraybackslash}p{2.5cm} 
  c 
  c 
  c 
} 
	\toprule
 		\textbf{Task} & 		\textbf{Participant} & 		\textbf{Session} & 		\textbf{Semantic Material} & 		\textbf{Chapter} & 		\textbf{Number of Characters} & 		\textbf{Duration}  \\\midrule 
\multirow{4}{*}{RA} & \multirow{2}{*}{f1, f2, m2} & 1 & The Little Prince & 1-27 & 21829 & 1h32min42s \\
    & & 2 & GarnettDream & 1-9 & 21512 & 1h29min23s \\ 
\cmidrule(lr){2-7} 
    & \multirow{2}{*}{m1} & 1 & The Little Prince & 1-27 & 21829 & 1h32min42s \\
    & & 2 & GarnettDream & 1-9 & 24,762 & 1h42min51s \\ 
\midrule
\multirow{16}{*}{PL} & \multirow{8}{*}{01-04} & \multirow{2}{*}{1} & The Little Prince & 1-7 & 5762 & 24min43s \\
    & & & GarnettDream & 1-3 & 6465 & 27min11s \\ 
    & & \multirow{2}{*}{2} & The Little Prince & 8-14 & 5904 & 25min16s \\
    & & & GarnettDream & 4-5 & 6312 & 26min30s \\ 
    & & \multirow{2}{*}{3} & The Little Prince & 15-24 & 6224 & 26min50s \\
    & & & GarnettDream & 6-7 & 4529 & 19min2s \\
    & & \multirow{2}{*}{4} & The Little Prince & 25-27 & 3746 & 15min52s \\
    & & & GarnettDream & 8-9 & 3963 & 16min41s \\ 
\cmidrule(lr){2-7} 
    & \multirow{8}{*}{05-08} & \multirow{2}{*}{1} & The Little Prince & 1-7 & 5762 & 24min43s \\
    & & & GarnettDream & 1-3 & 9844 & 41min16s \\ 
    & & \multirow{2}{*}{2} & The Little Prince & 8-14 & 5904 & 25min16s \\
    & & & GarnettDream & 4 & 4880 & 20min26s \\ 
    & & \multirow{2}{*}{3} & The Little Prince & 15-24 & 6224 & 26min50s \\
    & & & GarnettDream & 5-6 & 5016 & 21min4s \\ 
    & & \multirow{2}{*}{4} & The Little Prince & 25-27 & 3746 & 15min52s \\
    & & & GarnettDream & 7-8 & 4791 & 20min5s \\ 
\bottomrule
\end{tabular}
\end{table}

\begin{table}[h]
\centering

\begin{tabular}{l l} %
\hline
\textbf{Trigger} & \textbf{Description}  \\ \hline
BEGN & Start of EEG data collection by the EGI device \\ \hline
STOP & Stop collecting EEG data \\ \hline
CHxx & Start of each chapter, where xx is the chapter number (e.g., the first chapter is CH01) \\ \hline
ROWS & Start of a new line of text \\ \hline
ROWE & End of a line \\ \hline
PRES & Start of the preface reading phase \\ \hline
PREE & End of the preface reading phase \\
 \hline
\end{tabular}

\end{table}

\begin{figure}[ht]
\centering
\includegraphics[width=\linewidth]{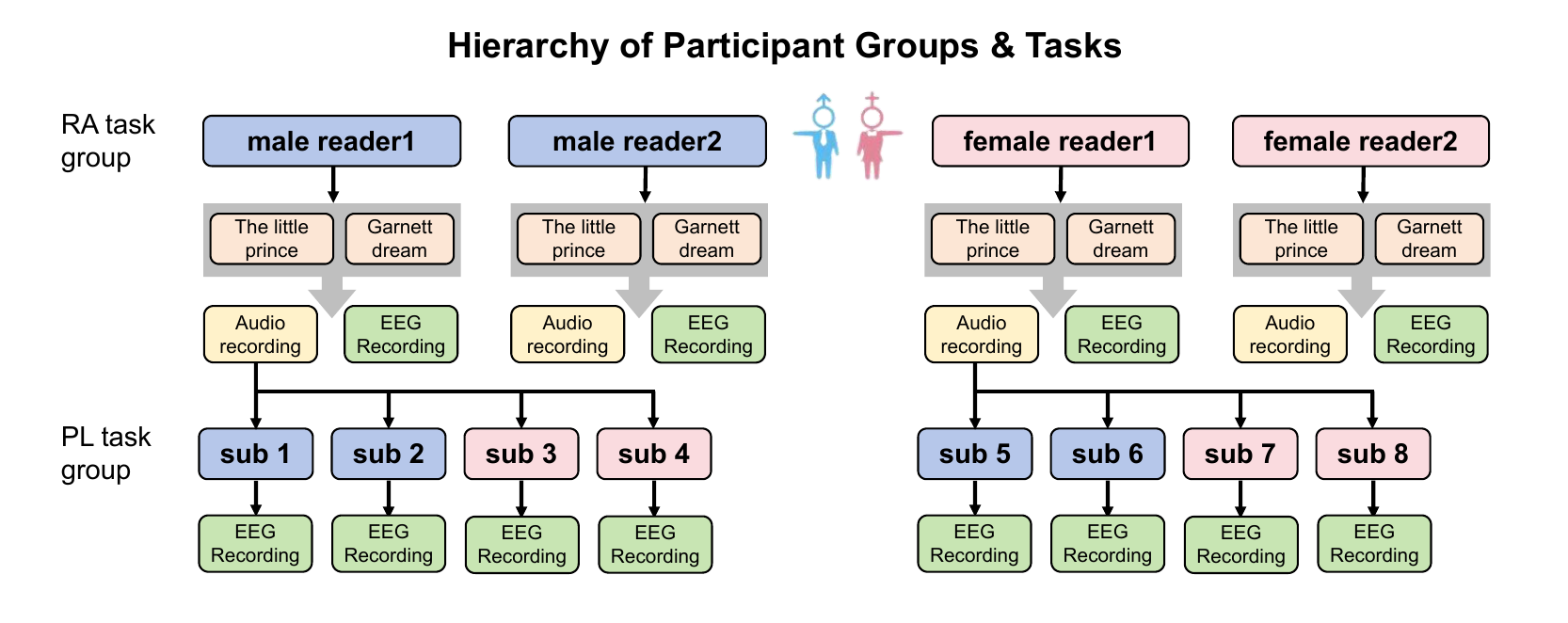}
\caption{\textcolor{red}{Overview of the experimental pipeline followed by the participants. During the reading phase, four participants independently read two novels while their EEG signals were recorded. In the listening phase, eight participants were divided into two groups (four participants per group), and each group listened to the novel audio recordings by Female Participant 1 and Male Participant 1 while their EEG activity was recorded.} }
\label{fig:participants}
\end{figure}


\end{document}